\documentclass[showpacs,prb,twocolumn,preprintnumbers,amsmath,amssymb]{revtex4}

\usepackage{graphicx}
\usepackage{dcolumn}
\usepackage{bm}

\begin{document}


\title{Iron self-diffusion in FeZr/$^{57}$FeZr multilayers measured by neutron reflectometry:\\
 Effect of applied compressive stress}
\author{Mukul Gupta $^{1,2}$}\email{mgupta@csr.ernet.in}\author{Ajay Gupta$^{1}$, Sujoy
Chakravarty$^{1}$, Rachana Gupta$^{1,2}$ and Thomas
Gutberlet$^{2}$}

\affiliation{$^{1}$UGC-DAE Consortium for Scientific Research,
Khandwa Road, Indore, 452017, India\\$^{2}$Laboratory for Neutron
Scattering, ETH Z\"{u}rich and Paul Scherrer Institute, CH-5232
Villigen PSI, Switzerland}

\date{\today}

\begin{abstract}
Iron self-diffusion in nano-composite FeZr alloy has been
investigated using neutron reflectometry technique as a function
of applied compressive stress. A composite target of Fe$+$Zr and
$^{57}$Fe$+$Zr was alternatively sputtered to deposit chemically
homogeneous multilayer (CHM) structure,
[$^{natrural}$Fe$_{75}$Zr$_{25}$/$^{57}$Fe$_{75}$Zr$_{25}$]$_{10}$.
The multilayers were deposited on to a bent Si wafer using a
3-point bending device. Post-deposition, the bending of the
substrate was released which results in an applied compressive
stress on to the multilayer. In the as-deposited state, the alloy
multilayer forms an amorphous phase, which crystallizes into a
nano-composite phase when heated at 373 K. Bragg peaks due to
isotopic contrast were observed from CHM, when measured by neutron
reflectivity, while x-ray reflectivity showed a pattern
corresponding to a single layer. Self-diffusion of iron was
measured with the decay of the intensities at the Bragg peaks in
the neutron reflectivity pattern after thermal annealing at
different temperatures. It was found that the self-diffusion of
iron slows down with an increase in the strength of applied
compressive stress.
\end{abstract}

\maketitle

\section{\label{Introduction}Introduction}

During recent decades amorphous and nanocrystalline metals and
alloys have been investigated as an important class of materials
with the possibility of tailoring their properties over a wide
range by controlling particle size and
morphology.~\cite{Debenedetti_Nature01,Angell_JAP00,McHenry_PMS99,Dunlop_PRL03}
More recently nano-composite alloys, in which nanocrystals are
surrounded by an intergranular amorphous matrix, have attracted a
great attention due to their interesting
structural~\cite{Ma_NatureMat02} and magnetic
properties.~\cite{Kojima_JAP00,Kojima_MSE97,Zhang_PRB02} One of
the favorable way to obtain a nano-composite alloy, is partial
crystallization of the amorphous
alloy.\cite{Angell_JAP00,Kappes_Nano03,Botta_JNCS99} The alloy
structure obtained above the primary crystallization temperature
of the parent amorphous phase, but below the secondary
crystallization temperature, has been termed as nano-composite
phase.~\cite{Hono_MatChar00} Above the secondary crystallization
temperature, the nano-composite structure fully crystallizes to
form an equilibrium state of the alloy. Structurally, the
nanocrystals obtained after primary crystallization are surrounded
by an amorphous inter-granular phase to form a nano-composite
phase.~\cite{Debenedetti_Nature01,Zhu_JP04} Nano-composite alloys
produced with an amorphous precursor are the basis of interesting
soft-magnetic alloys known as FINEMET~\cite{Yoshizawa_JAP88},
NANOPERM~\cite{Suzuki_MatTrans91}, HITPERM~\cite{Willard_JAP99}.
Since the nano-composite phases produced in these alloys is
inherently a metastable phase, diffusion of the constituents would
play an important role in understanding and determining their
properties for long-standing applications. Hence, atomic diffusion
in such alloys is the key phenomenon for selecting their
applications.~\cite{Fauple_RMP03}

The situation becomes more complicated when the nano-composite
alloys are produced in the form of a thin film. Deposition of thin
films on to a substrate is known to produce films with a large
intrinsic strain or stresses which often results from differences
in thermal expansion (thermal stress) or from the microstructure
of the deposited film (intrinsic
stress).~\cite{Abermann_TSF85,Shull_JAP96,Floro_JAP01} The
intrinsic stresses may originate due to several factors (i) at the
strained regions within the films e.g. grain-boundaries,
dislocations, voids, impurities, etc. (ii) at the film/substrate
interface due to lattice mismatch, different thermal expansion,
etc. (iii) at the film/vacuum interfaces due to surface stress,
adsorption, etc. or (iv) due to a dynamic processes e.g.
re-crystallization, interdiffusion, etc.~\cite{Koch_JPCM94}  These
stresses may significantly affect the physical properties of the
thin films, including atomic diffusion.

It is known that when a material is deposited in the thin film
state, the diffusion mechanism can be completely different as
compared to bulk state of that material, even when the material is
in purely elemental form. Such a behavior has been mainly
attributed to an increased defect concentration, metastability and
un-relaxed state of the material. Therefore an extrapolation of
bulk diffusivity may results in erroneous values of diffusivity in
the case of thin films. Since many devices which are used for
application are fabricated in the form of nm range thin films,
self-diffusion measurements can be extremely important for their
applications.

In order to study the nature of stresses on self-diffusion we have
chosen a simple binary FeZr alloy for this purpose. It was found
that (as will be shown later) after annealing at 373 K, the alloy
forms a nano-composite phase which further crystallizes above 600
K. The self-diffusion of iron was measured in the nano-composite
state as a function of applied stress. The samples were deposited
on to a substrate with a known bending. An external stress on to
an
[$^{natrural}$Fe$_{75}$Zr$_{25}$/$^{57}$Fe$_{75}$Zr$_{25}$]$_{10}$
multilayer was applied by releasing the bending of the substrate
which resulted in an applied compressive stress on to the sample.
Iron self-diffusion measurements were carried out using neutron
reflectivity technique. It may be noted that the neutron
reflectivity is an excellent technique for studying self-diffusion
in nm range structures. Due to the fact that neutron scattering
length densities for isotopes of an element are different, neutron
reflectivity with depth resolution in sub nm range provides a
unique opportunity for measuring self-diffusion. Conventional
cross-sectioning and depth-profiling techniques, such as
radiotracer, secondary ion mass spectroscopy (SIMS) are not
suitable for measuring self-diffusion in nm range structures as
the depth resolution available with cross-sectioning and
depth-profiling techniques is of the order of a few nm.

In an earlier work~\cite{Gupta_PRB04} we have demonstrated that
neutron reflectivity is a technique which could be used to probe
diffusion lengths of the order of 0.1 nm, and diffusion at
temperatures less than 400 K could be measured. In the present
work, the effect of compressive stress on the self-diffusion of
iron in nano-composite multilayers was studied using neutron
reflectivity.

\section{\label{Char}Sample preparation and characterization}

FeZr, CHM have been deposited on Si (100) substrates using
magnetron sputtering technique. Small pieces of Zr rods were
pasted on the $^{natural}$Fe and $^{57}$Fe targets in a symmetric
way and the composite targets were sputtered alternately to
prepare a chemically homogeneous structure with nominal a
composition Si/[$^{natrural}$Fe$_{75}$Zr$_{25}$ (25
nm)/$^{57}$Fe$_{75}$Zr$_{25}$(10 nm)]$_{10}$. The deposition of
the multilayer was carried out after obtaining a base pressure
better than 1$\times$10$^{-6}$ mbar. During the deposition,
pressure in the chamber was 5$\times$10$^{-3}$ mbar due to 30
cm$^{3}$/min Ar gas flow used for sputtering of the targets. All
the samples were deposited at a constant sputtering power of 50 W.
Before deposition the vacuum chamber was repeatedly flushed with
Ar gas so as to minimize the contamination of the remaining gases
present in the chamber. Both the targets were pre-sputtered at
least for 10 min. During the deposition the substrate was mounted
on a specially designed 3-point Si wafer bending device. The
substrate was oscillated with respect to central position of the
target for better uniformity of the thickness of the deposited
sample.

\begin{figure} \vspace{-25mm}
\includegraphics [width=80mm,height=80mm] {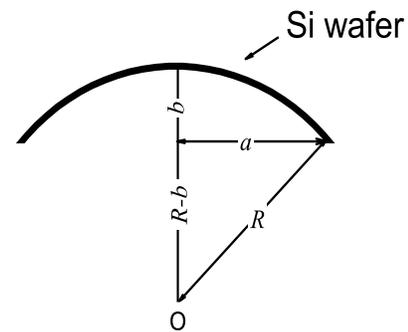} \vspace{-5mm}
\caption{\label{fig:Fig1} Schematic diagram of the bent Si wafer,
used for calculation of radius of curvature.}
\end{figure}

In all the cases, thin Si wafers (300$\pm$10) $\mu$m were used as
a substrate in order to avoid breaking during bending. The Si
wafer was fixed from both the ends, and by rotation of an
asymmetric roller around the central axis, the bending height of
the Si wafer can be varied between 0 to 5 mm. A pin-lock system
was incorporated so that release of bending by itself could be
avoided. A compressive stress is applied on to the deposited film,
when the bending of the Si wafer was released after deposition.
The applied stress due to release of bending on to the Si wafer
can be calculated using Stoney's formula~\cite{Stoney_PRS1909} and
following a discussion given by Chen et al.~\cite{Chen_SST03} The
applied stress $\sigma$, is given by:

\begin{equation}
\sigma = \frac{(\frac{E_{Si}}{1-\nu_{Si}})T_{Si}^{2}}{6RT_{f}},
\label{eq:one}
\end{equation}

Where $(\frac{E_{Si}}{1-\nu_{Si}})$ is the biaxial modulus of the
silicon substrate and is equal to 180.5 GPa. E$_{Si}$ is Young's
modulus for Si and $\nu_{Si}$ is Poisson's ratio for Si. T$_{Si}$
is the thickness of the substrate, T$_{f}$ is the thickness of the
film and $R$ is the radius of curvature. With the situation shown
in the fig.~\ref{fig:Fig1}, the radius of curvature can be written
as:

\begin{equation}
R = \frac{a^{2}+b^{2}}{2b}, \label{eq:two}
\end{equation}

Combining equation~(\ref{eq:one}) and~(\ref{fig:Fig2:XRR Asd 0 27
46 GPa}), the value of stress was calculated. The parameter used
in the present case are T$_{Si}$ = (300$\pm$10)$\mu$m, T$_{f}$ =
370 nm, $a$ = 40 mm and $b$ was varied at 0, 3 and 5 mm. The
obtained values of stress for the 3 cases are 0, 27 and 46 GPa.
The errors in the calculation of applied stress were of the order
of 15-20\%, taking into account the uncertainties in the measured
physical parameters. Samples with different known bending were
deposited under similar deposition conditions. After deposition
and release of bending, the surface profile of the samples was
measured using a profilometer. It was found that the surface of
the samples was flat and no changes in the surface profile were
observed for a sample prepared with or without bending. This
indicated that even after the bending the substrate gains its
original state and the stress is applied on to the deposited
multilayer.

The composition of the deposited films was determined using x-ray
photoelectron spectroscopy (XPS) depth profiling. The XPS profile
was measured using monochromatic Al K$_{\alpha}$ x-rays (1mm spot
size) at the surface and at 3 different depths, after sputtering
with Ar ions of 1 mA current and 3 kV accelerating voltage. The
pressure during measurements was better than 1$\times$10$^{-8}$
mbar. The average composition of the films was equal to
Fe$_{75\pm3}$Zr$_{25\pm3}$, excluding the data taken at the
surface. Since at the surface, contributions from absorbed species
like carbon and oxygen were significant, the average composition
of the film was determined with the data taken at 3 different
depths. A small amount of oxygen was detected throughout the depth
of the film.

Structural characterizations of the samples were carried out with
x-ray reflectivity (XRR) and grazing incidence diffraction using a
standard x-ray diffractometer (XRD) with Cu-K$_{\alpha}$ x-rays.
The crystallization behavior of the multilayers was examined using
differential scanning calorimetry (DSC) with NETZSCH, DSC equipped
with extremely high sensitivity $\mu$-Sensor.  The conversion
electron M\"{o}ssbauer spectroscopy measurements (CEMS) were
performed for determining the local environment of $^{57}$Fe in
the samples. The measurements were carried out using a 50 mCi
$^{57}$Co-radiactive source in a Rh matrix and a gas flow
proportional counter (He+4\%CH$_{4}$) for detection of conversion
electrons. The isomer shifts were calibrated relative to
$\alpha$-Fe. Hysteresis loops as a function of azimuthal angle
were measured using magneto optical Kerr effect (MOKE).

Self-diffusion measurements were performed using neutron
reflectometry technique at AMOR reflectometer at the Swiss
spallation neutron source (SINQ), at Paul Scherrer Institute,
Switzerland.~\cite{Gupta_PJP04} The reflectivity pattern was
measured using two different angular settings (0.5$^{\circ}$ and
1.0$^{\circ}$) in the time-of-flight mode.

\section {{\label{sec:RnD}}Results and Discussion}

\subsection {{\label{sec:RnD:StrucProp}}Structural Properties}

\begin{figure}\vspace{-5mm}
\includegraphics[width=80mm,height=70mm]{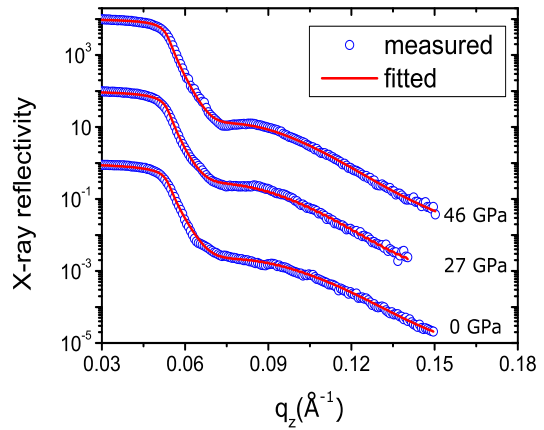} \vspace{-5mm}
\caption{\label{fig:Fig2:XRR Asd 0 27 46 GPa} (color online) X-ray
reflectivity of the as-deposited
Si/[$^{natrural}$Fe$_{75}$Zr$_{25}$ (25$\pm$1
nm)/$^{57}$Fe$_{75}$Zr$_{25}$(12$\pm$1 nm)]$_{10}$ multilayer at
different applied stresses. The intensity shown on y-axis has been
multiplied by a factor of 100, for clarity.}
\end{figure}

The multilayers prepared in this work have a periodicity only for
iron isotopes, it is expected that x-ray reflectivity of the
multilayers would show a pattern corresponding to a single layer.
Fig.~\ref{fig:Fig2:XRR Asd 0 27 46 GPa} shows x-ray reflectivity
pattern of the multilayer structure prepared at 0, 27 and 46 GPa.
As can be seen from the pattern, at the designed period of the
multilayer there was no contrast for x-rays, which confirms the
chemical homogeneity of the layers. The x-ray reflectivity pattern
was fitted assuming a single layer and an `oxide' layer of about 6
nm thickness on the surface of the multilayer, using a computer
program based on Parratt's formalism.~\cite{Parratt_PR54} Such an
oxide layer on the surface of the sample may stem because of
absorbed oxygen or other light elements, when the samples are
exposed to the atmosphere after deposition. Presence of such a
layer was also evident from XPS measurements.

On the other hand the neutron reflectivity pattern
(fig.~\ref{fig:Fig3: NR Asd 0, 27 and 46 GPa}), showed
well-pronounced Bragg peaks arising due to isotopic contrast
between $^{natural}$Fe and $^{57}$Fe. As can be seen from the
figure, the sample prepared without any stress showed rather
asymmetric Bragg peaks, while for the samples prepared with an
applied stress, the peaks were more symmetric. Such an asymmetry
of the Bragg peaks may arise due to incorporation of some free
volume which may result in some internal strain or stresses during
the growth of the film and might cause an asymmetry in the
scattering length density. For the samples which were prepared in
the bent state, the release of bending results into an applied
external stress on to the multilayer which eventually results in
annihilation of free volume. A more detailed discussion related to
this issue is given in the next sections.

\begin{figure} \vspace{-10mm}
\includegraphics[width=80mm,height=70mm]{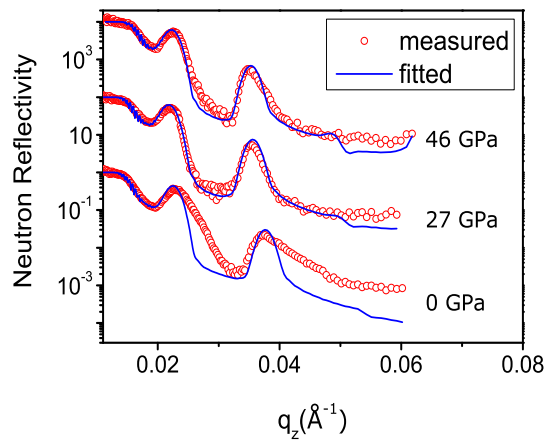} \vspace{-5mm}
\caption{\label{fig:Fig3: NR Asd 0, 27 and 46 GPa} (color online)
Neutron reflectivity of the as-deposited
Si/[$^{natrural}$Fe$_{75}$Zr$_{25}$ (25$\pm$1
nm)/$^{57}$Fe$_{75}$Zr$_{25}$(12$\pm$1 nm)]$_{10}$ multilayer
prepared with applied stresses of 0, 27 and 46 GPa. The intensity
shown on y-axis has been multiplied by a factor of 100, for
clarity. }
\end{figure}

The neutron reflectivity pattern was fitted using a computer
program based on Parratt's formalism~\cite{Parratt_PR54} and it
was found that the pattern could not be fitted assuming sharp
interfaces; instead a thin inter-layer of thickness (1$\pm$0.5) nm
with the mean scattering length density of the two layers had to
be introduced as inter-diffused layer. This means that already at
room temperature there is some amount of interdiffusion in the
multilayer. The fitted parameters gives the structure of the
multilayers: Si/[$^{natrural}$Fe$_{75}$Zr$_{25}$ (25$\pm$1
nm)/$^{57}$Fe$_{75}$Zr$_{25}$(12$\pm$1 nm)]$_{10}$, which is close
to the nominal structure.

\subsection {{\label{sec:RnD:CrystBehav}}Crystallization Behavior}

\begin{figure}
\includegraphics[width=80mm,height=70mm]{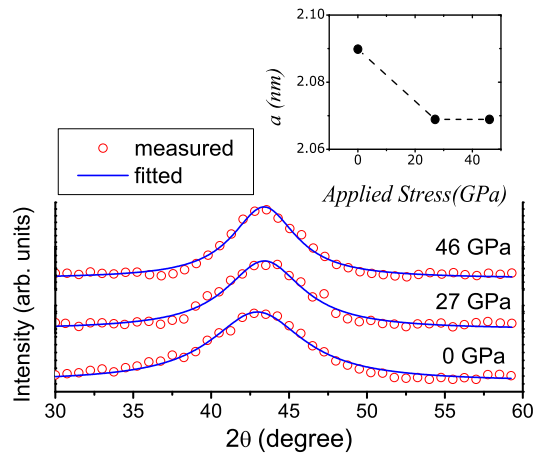}
\caption{\label{fig:Fig4} (color online) X-ray diffraction pattern
of the isotopic multilayers in the as-deposited state. The
measurements were carried out in the grazing incidence geometry
using Cu-K$\alpha$ x-rays. The inset in the figure shows the
change in inter-atomic distance as a function of annealing
temperature.}
\end{figure}

\begin{figure}
\includegraphics[width=80mm,height=100mm]{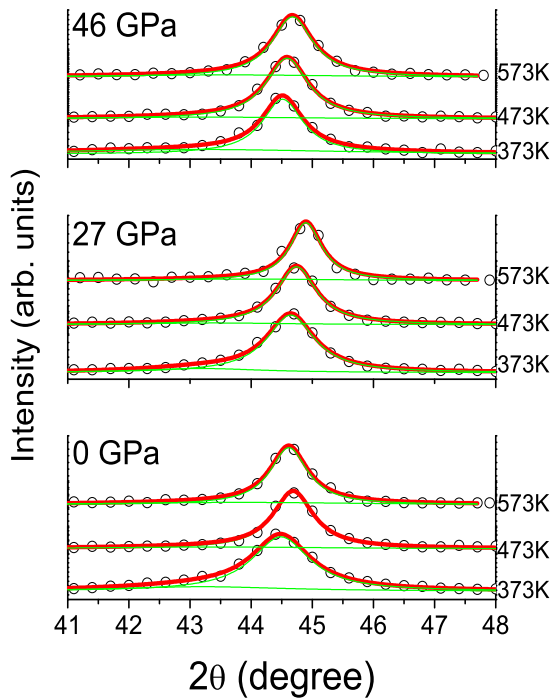}
\caption{\label{fig:Fig5} (color online) Grazing incidence x-ray
diffraction pattern of Si/[$^{natrural}$Fe$_{75}$Zr$_{25}$
(25$\pm$1 nm)/$^{57}$Fe$_{75}$Zr$_{25}$(12$\pm$1 nm)]$_{10}$
multilayer prepared with an applied stress of 0, 27 and 46 GPa
after annealing at different temperatures. Open circles represents
the measured data and the solid lines are fit to them.}
\vspace{-5mm}
\end{figure}

\begin{figure}\vspace{-5mm}
\includegraphics[width=70mm,height=60mm]{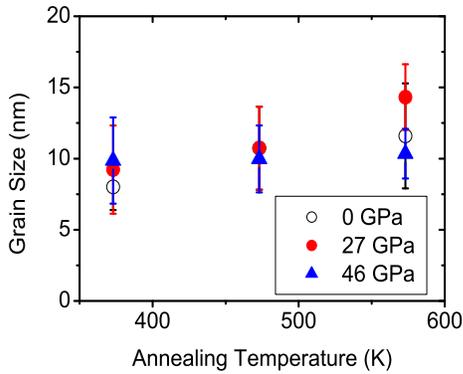}
\caption{\label{fig:Fig6} (color online) Average grain size as a
function of annealing temperature shown representatively for a
sample prepared with an applied stress of 0, 27 and 46 GPa.}
\vspace{-5mm}
\end{figure}

Prior to diffusion measurements thermal stability of the samples
was studied with grazing incidence x-ray diffraction. All the
samples were annealed together in a vacuum furnace in the
temperature range of 373-573 K with 100 K step. In the as-prepared
state all the samples show a diffuse maxima cantered around
2$\theta$ = 44.6$^{\circ}$, with a width of about 4-5$^{\circ}$
(see Fig.~\ref{fig:Fig4}), which means that the samples are x-ray
amorphous in the as-prepared state. The width of the diffuse
maxima is comparable to the iron based amorphous
alloys.~\cite{Angell_JAP00} The average interatomic distance can
be estimated using the relation,~\cite{Guinier_XRD}
$a=1.23\lambda/2\sin\theta$, where $\theta$ is taken to be the
angle at the center of the diffuse maxima, and the factor 1.23 is
a geometric factor which rationalizes the nearest neighbor
distance with the spacing between `pseudo-close packed planes'. As
shown in the fig.~\ref{fig:Fig4}, with an increase in the applied
stress the position of the amorphous maxima shifts towards higher
angle side indicating a decrease in the average inter-atomic
distance as shown in the inset of the figure. Such a decrease in
the average inter-atomic distance may be caused due to applied
compressive stress. After annealing at 373 K, the broad hump
becomes narrow (width $\sim$1$^{\circ}$) and a nano-composite
structure is found, as shown in fig.~\ref{fig:Fig5}. The peak
shape from the nano-composite structure could be fitted only by
deconvoluting it into two lines, one corresponding to the parent
amorphous phase and the second to a nanocrystalline bcc-Fe phase.
The area ratio of amorphous phase as determined from the fitting
of XRD data was in the range of 15-25\%. A slight decrease in the
area ratio of the amorphous phase was observed. On further
annealing at 473 and 573 K, no significant changes in the XRD
pattern of the samples were observed as shown in
fig.~\ref{fig:Fig5}. After annealing at different temperatures the
position of Bragg peak shifts towards higher angle indicating a
further decrease in the interatomic spacing. Such a decrease in
interatomic spacing is related with structural relaxation and is a
consequence of annihilation of free volume. The grain size of the
nanocrystals was about 10 nm, which increases marginally with an
increase in the annealing temperature as shown in
fig.~\ref{fig:Fig6}.

\begin{figure}
\includegraphics[width=80mm,height=110mm]{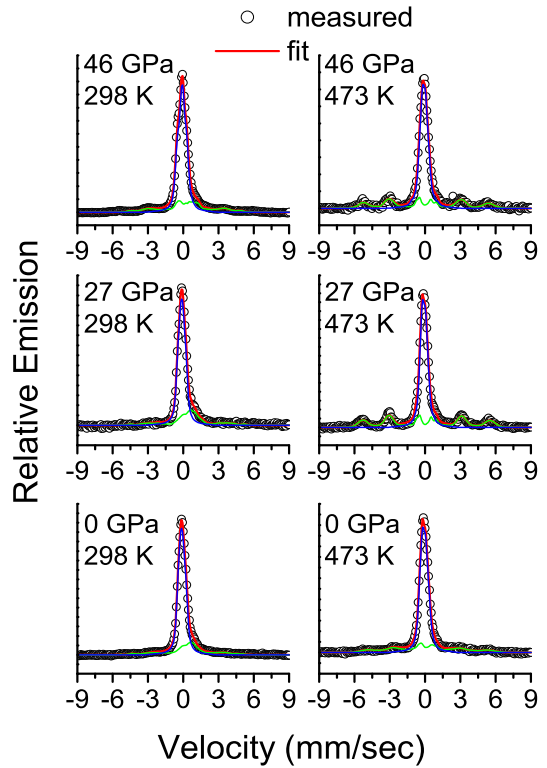}
\caption{\label{fig:Fig7} (color online) Conversion electron
M\"{o}ssbauer spectroscopy (CEMS) pattern of
Si/[$^{natrural}$Fe$_{75}$Zr$_{25}$ (25$\pm$1
nm)/$^{57}$Fe$_{75}$Zr$_{25}$(12$\pm$1 nm)]$_{10}$ multilayer
prepared with an applied stress of 0, 27 and 46 GPa in the
as-deposited state and after annealing at 473 K.}
\end{figure}

Crystallization behavior of the samples was also studied using
conversion electron M\"{o}ssbauer spectroscopy (CEMS). Samples
prepared with and without applied stress were annealed at high
temperatures. Fig.~\ref{fig:Fig7} compares CEMS pattern of a
sample prepared with and without applied stress before and after
annealing at 473 K. The CEMS patterns were fitted assuming a
sextate due to Fe nanocrystals and a doublet due to amorphous
phase. Even in the as-deposited state a sextate with hyperfine
field of about 10 Tesla was found to be present. However, the area
ratio of this sextate is very small. This indicates that there is
a small amount of magnetic alloy while most of the alloy is not
ferromagnetic. As the samples were annealed the contribution of
this magnetic phase increases indicating an enhancement in the
volume fraction of nanocrystalline Fe in agreement with the XRD
results. It may be noted that the hyperfine field after annealing
remains in the range of 18-23 Tesla while that of pure Fe is 33.3
Tesla. The reduced magnetic moment could result due to some
thermal fluctuations. Table ~\ref{Tab:Table1} compares the fitted
values for the different cases as shown in fig.~\ref{fig:Fig7}.

\begin{table} [!!h]
\caption{\label{Tab:Table1}Fitted CEMS parameters for the sample
prepared with and without an applied stress in the as-deposited
state and after annealing at 473 K.}

\begin{ruledtabular}
\begin{tabular}{cccc}
 Sample    & Sample      & Average    & Average  \\
          & Condition & Hyperfine &Quadrupole\\
          & & Field (T) & (mm s$^{-1}$) \\

\hline
 & & & \\
0 GPa     & As-deposited         & 10.8$\pm$0.5 T            & 0.40$\pm$0.01  \\
0 Gpa     & 473 K, 1 hour        & 18.4$\pm$1.3 T            & 0.43$\pm$0.01 \\
27 GPa    & As-deposited         & 10.2$\pm$0.9 T            & 0.44$\pm$0.01 \\
27 Gpa    & 473 K, 1 hour        & 23.4$\pm$0.8 T            & 0.70$\pm$0.01 \\
46 GPa    & As-deposited         & 13.0$\pm$0.2 T            & 0.44$\pm$0.01 \\
46 Gpa    & 473 K, 1 hour        & 22.5$\pm$1.5 T            & 0.42$\pm$0.01 \\
\end{tabular}
\end{ruledtabular}
\end{table}

\begin{figure*}

\includegraphics [width=145.8mm,height=60mm] {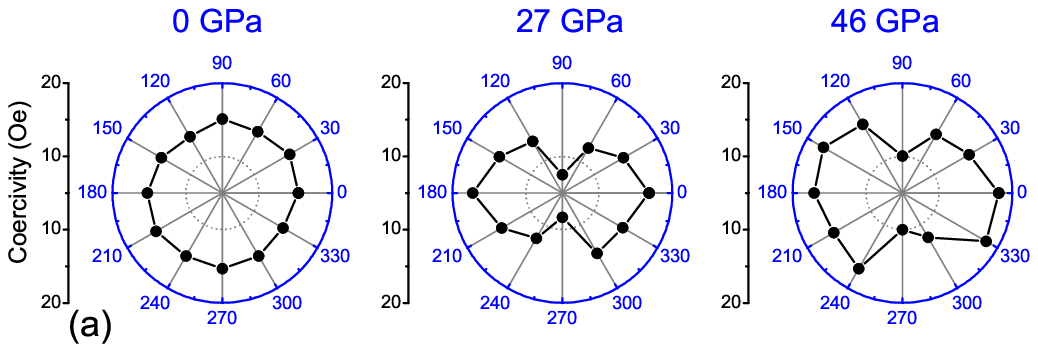} \vspace{-5 mm}
\includegraphics [width=145.8mm,height=60mm]{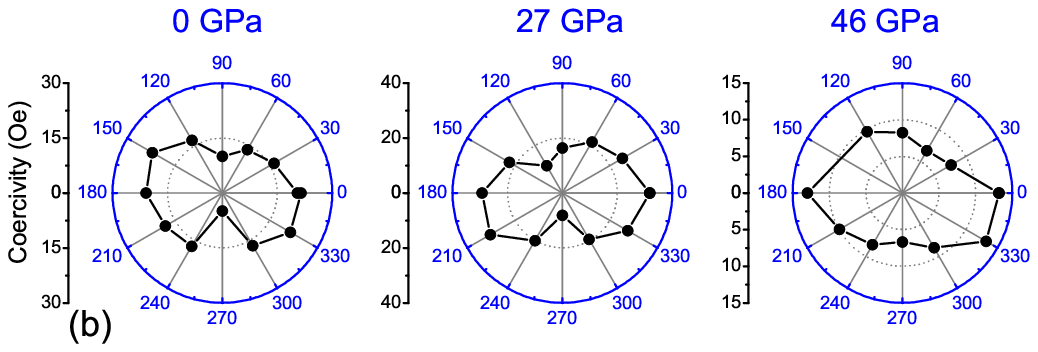}
\caption{\label{fig:Fig8} (Color online) MOKE measurements on
Si/[$^{natrural}$Fe$_{75}$Zr$_{25}$ (25$\pm$1
nm)/$^{57}$Fe$_{75}$Zr$_{25}$(12$\pm$1 nm)]$_{10}$ multilayer
prepared with an applied stress of 0, 27 and 46 GPa in the
as-deposited state (a) and after annealing at 473 K (b).}
\end{figure*}

MOKE measurements were also performed to understand the changes in
the magnetic properties of the system during amorphous to
nanocomposite phases. The sample prepared without an applied
stress exhibited no anisotropy as a function of azimuthal angle in
the coercivity. While the samples prepared with an applied stress
clearly show uniaxial anisotropy, which persists even after the
nanocrystallization of the films (see fig.~\ref{fig:Fig8}). This
confirms that the bending stress induced in the films persists
even after nanocrystallization.

Formation of a nano-composite phase after primary crystallization
of the amorphous phase is a general phenomenon in amorphous
alloys. Often, it was observed that amorphous binary alloys
crystallize in two steps. The primary crystallization reaction of
most of amorphous alloys leads to an evolution of nanocrystalline
microstructures whereas the phase formed after the second stage
results in an intermetallic compound along with the
nanocrystalline phase. The nominal reaction for such
crystallization process had been given as: amorphous $\rightarrow$
$\alpha$+amorphous $\rightarrow$ $\alpha+\beta$; where $\alpha$ is
the primary phase that precipitates out from the amorphous matrix
and $\beta$ is an intermetallic
compound.~\cite{Angell_JAP00,Zhu_JP04,Hono_MatChar00} In the
present case crystallization of the amorphous phase can be
regarded as the primary crystallization process and as evident
from the x-ray data, the amorphous phase co-exists along with
grains of $\alpha$-Fe. However, the primary crystallization
temperature for the present case was found to be very low as
compared with Fe$_{67}$Zr$_{33}$ amorphous
alloy.~\cite{Gupta_PRB04} Since in the present case for
Fe$_{75}$Zr$_{25}$ alloy, the Zr content is slightly lower, a
decrease in crystallization temperature is not very surprising. In
order to further confirm the structure of the alloy, a thin film
with even lower Zr content was deposited under identical
conditions of sputtering.~\cite{Gupta_Unpub} The composition of
this film was Fe$_{80}$Zr$_{20}$. The XRD pattern of this film
showed a narrow peak even in the as-prepared state (not shown in
the figure), indicating that the structure forms a nanocrystalline
state. Such a decrease in the primary crystallization temperature
was also observed in an ion beam sputtered Fe$_{85}$Zr$_{15}$
sample,~\cite{Gupta_JNCS04,Gupta_PSS04} and a phenomenon analogue
to surface crystallization in amorphous alloy
ribbons~\cite{Gupta_HypInt90,Radlinski_PRL86,Rao_RSSSci89,Koster_MRSSP97}
was found responsible for early crystallization of amorphous
Fe$_{85}$Zr$_{15}$ film in the thin film state.

\begin{figure}
\includegraphics[width=80mm,height=70mm]{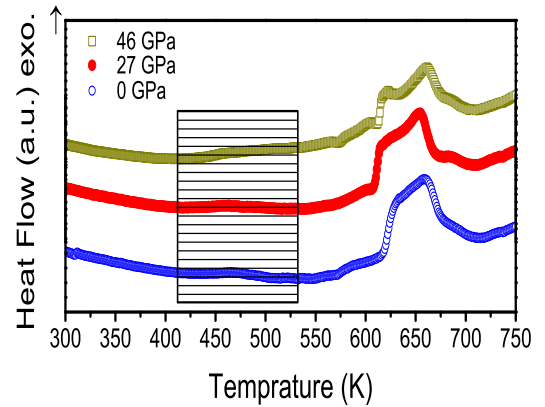}
\caption{\label{fig:Fig9} (color online) DCS scans of the
Si/[$^{natrural}$Fe$_{75}$Zr$_{25}$ (25$\pm$1
nm)/$^{57}$Fe$_{75}$Zr$_{25}$(12$\pm$1 nm)]$_{10}$ multilayer
prepared with an applied stress of 0, 27 and 46 GPa. The shaded
region shown in the figure correspondences to the temperature
range used for diffusion measurements.}
\end{figure}

For the amorphous thin film formed by vapor deposition, the
effective quenching rate is very high which results in a higher
quenched-in free volume and results in an early crystallization of
the amorphous phase as observed in the present case. In order to
further understand the crystallization behavior of the alloy, DSC
measurements were carried out under a constant heating rate of
0.33 K/s. It was found that a very broad hump appears around 450 K
and a relatively sharp peak appears around 613 K in all the three
samples. The hump appearing at 450 K can be estimated as first
crystallization step while relatively sharp peak corresponds to
second crystallization step. From the XRD results, the first
crystallization event was observed as early as 373 K, where the
samples were annealed for 1 hour, in the DSC scan since the
samples were heated at a much faster rate, the crystallization
event was observed at higher temperature as shown in
fig.~\ref{fig:Fig9}. In a number of studies performed on bulk or
thick films (thickness few $\mu$m), it has been found that the
heat release during the first crystallization event is
significantly smaller compared to the second crystallization event
due to a slower diffusion at lower
temperature.~\cite{Perepezko_JNCS03} In Al-based glasses, Foley et
al~\cite{Foley_SM96} have studied the crystallization behavior
using transmission electron microscopy (TEM) and DSC. While growth
of nanocrystals was confirmed by TEM, there was no evidence of
primary crystallization with DSC. Small diffusion of the
constituents was argued for the observed behavior. In their case,
they observed that, for diffusivity, D $\sim$ 1$\times$l0$^{-19}$
m$^{2}$s$^{-1}$, the level of heat output is nearly undetectable
in DSC measurements, which requires a signal on the order of 0.1
mW or greater. Unless the value of D was at least two orders of
magnitude larger, the signal will be close to the noise level of
the DSC. It may be noted that in the present case, the iron
self-diffusivity around 400 K, is of the order of 10$^{-21}$
m$^{2}$s$^{-1}$ (as shown in later sections). Further, the
grain-size as determined with XRD results was about 10 nm, the
heat release for the formation of small grain sizes is expected to
be small. In addition, the DSC measurements in the present case
were performed in relatively thin film (370 nm) and the total mass
exposed during DSC measurements was only 65 $\mu$gm which explains
small heat release during the first crystallization event, in
spite of the high sensitivity of the sensor used during DSC
measurements. The presence of peak around 613 K, can be understood
as second crystallization step. The onset of second
crystallization temperature was found around (608$\pm$5) K for all
the 3 samples and there was no systematic effect of applied stress
on the second crystallization temperature.


\subsection {{\label{sec:RnD:SelfDiff:Time}}Self-Diffusion Measurements - Time Dependence}

\begin{figure}
\includegraphics[width=70mm,height=60mm]{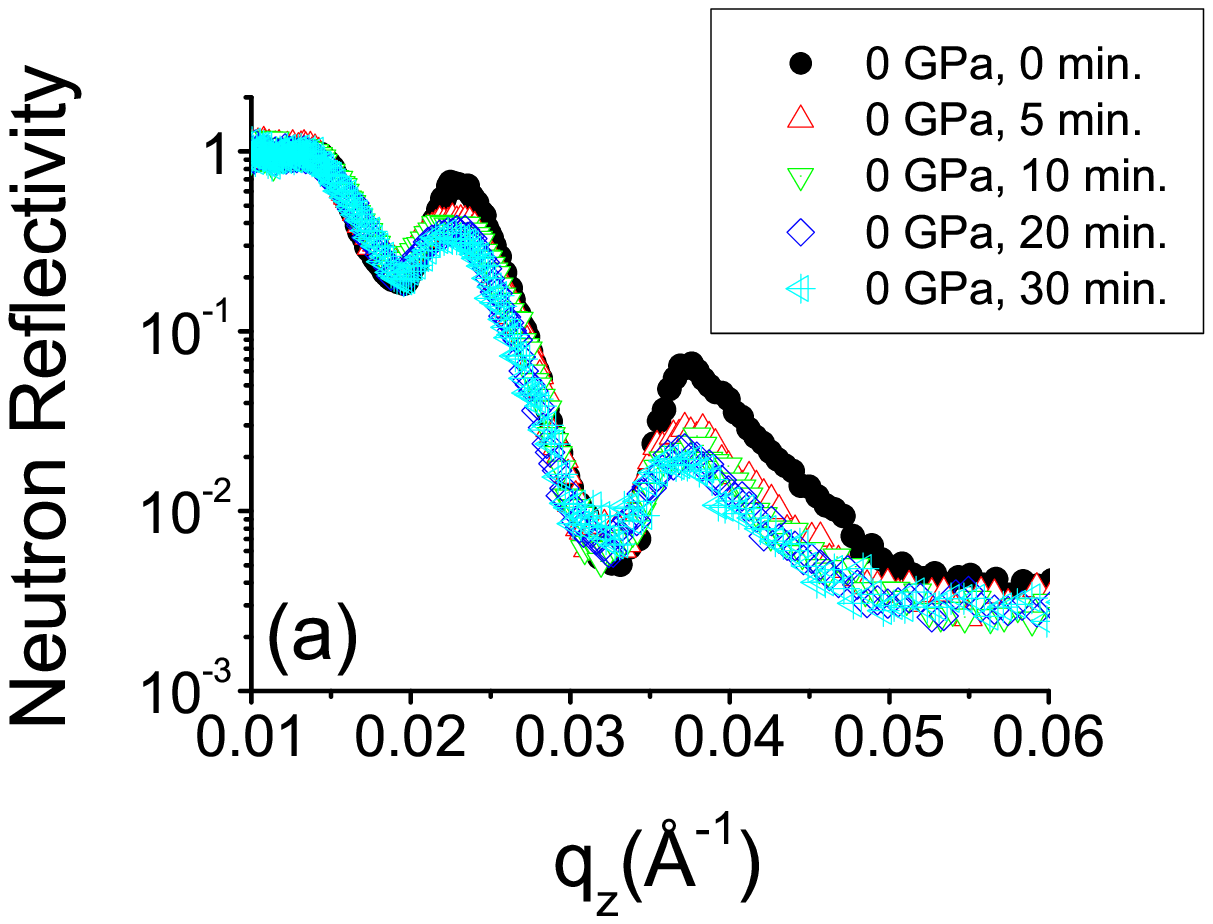} \vspace{-5mm}
\includegraphics[width=70mm,height=60mm]{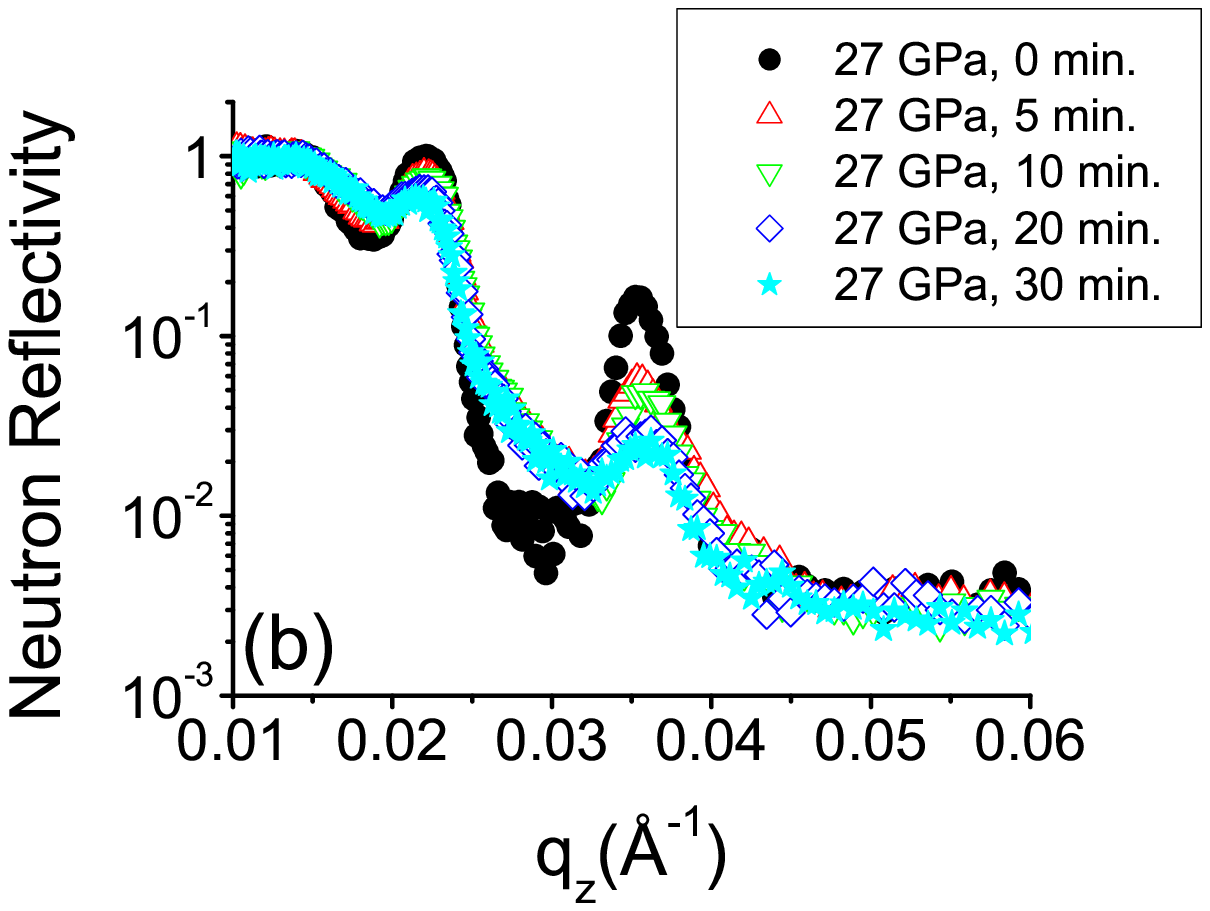} \vspace{-5mm}
\includegraphics[width=70mm,height=60mm]{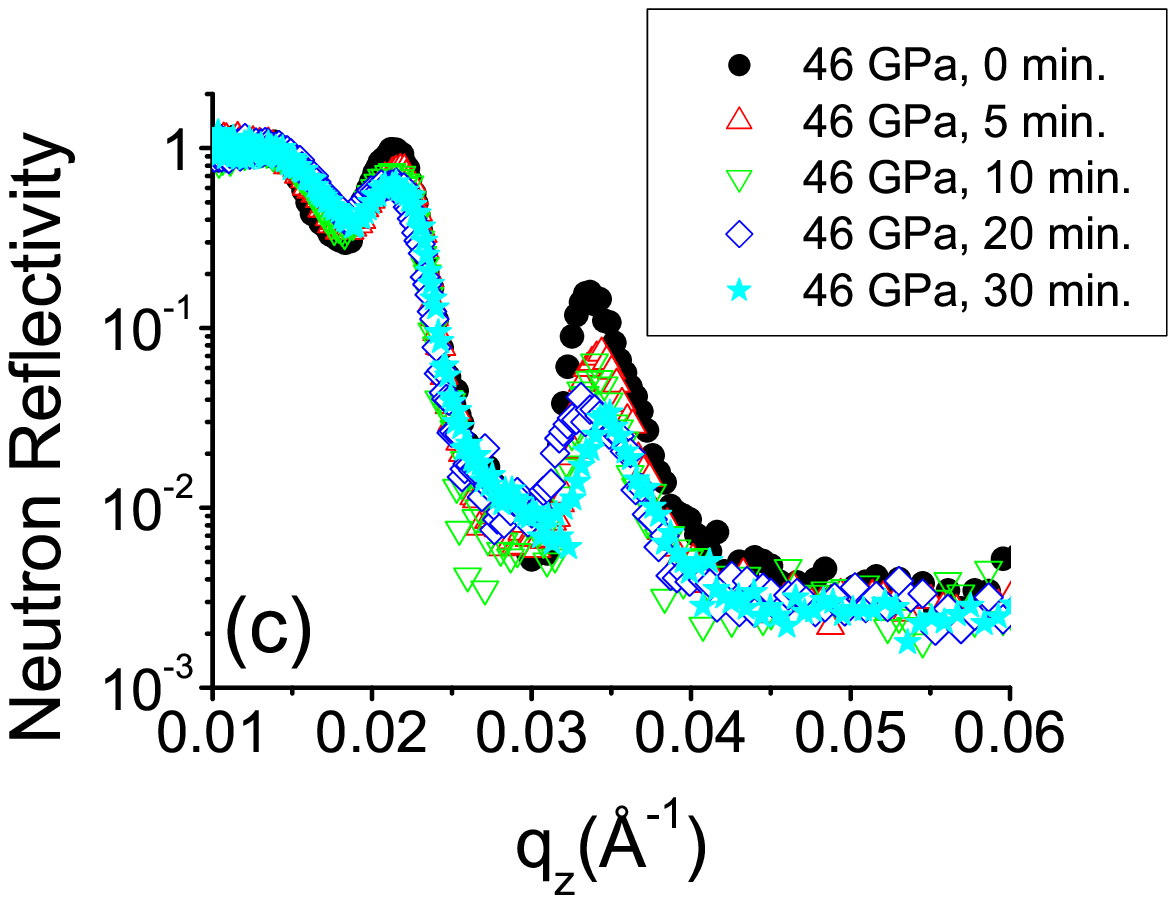}
\caption{\label{fig:Fig10} (color online) Decay of the Bragg peak
intensity as a function of annealing time at 473 K for the sample
prepared without an applied stress (a) with an applied stress of
27 GPa (b) and with an applied stress of 46 GPa (c).}
\vspace{-6mm}
\end{figure}

\begin{figure} \vspace{-5mm}
\includegraphics[width=70mm,height=60mm]{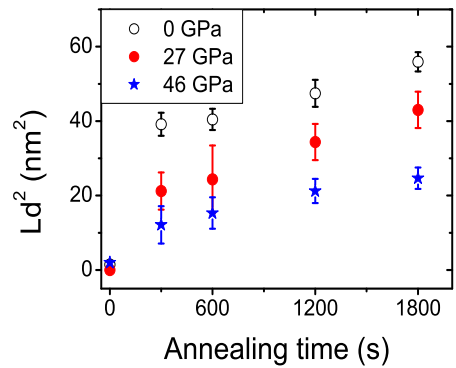}
\caption{\label{fig:Fig11} (color online)Evolution of diffusion
length as a function of annealing time and applied stress at 473 K
in nanocomposite Si/[$^{natrural}$Fe$_{75}$Zr$_{25}$(25$\pm$1
nm)/$^{57}$Fe$_{75}$Zr$_{25}$(12$\pm$1 nm)]$_{10}$.} \vspace{-5mm}
\end{figure}

With the observed thermal behavior of the samples, for diffusion
measurements, a temperature range for diffusion annealing was
chosen from 413-533 K in order to study diffusion in the
nano-composite state. This temperature region is also indicated in
fig.~\ref{fig:Fig8} as shaded area. The three samples prepared
with an applied compressive stress of 0, 27 and 46 GPa were first
pre-annealed at 373 K for 0.5 hour to obtain the nano-composite
phase. For studying the time dependence of diffusivity, the
samples were further annealed at 473 K and neutron reflectivity
measurements were carried out after each annealing. In order to
minimize the fluctuations due to a possible variation in the
temperature, all the samples were annealed simultaneously in the
furnace. Fig.~\ref{fig:Fig10} shows a plot of neutron
reflectivities as a function of annealing time at 473 K. A
relatively small time step was chosen in order to observe the
structural relaxation of the samples. As can be seen from the
figure, after annealing, the intensity at the Bragg peak decays.
The initial decay was found to be much faster as compared to that
with later annealing time. The decay of the Bragg peak intensity
can be used to calculate the diffusion coefficient using the
expression~\cite{Speakman_JMMM96}:

\begin{equation}
    ln [I(t)/I(0)] = -~8 \pi^{2} n^{2} D(T) t / d^{2},
\label{eq:three}
\end{equation}

where $I(0)$ is the intensity before annealing and $I(t)$ is the
intensity after annealing time $t$ at temperature T. The diffusion
length $L_{d}$ is related to diffusivity through the relation:

\begin{equation}
    L_{d} = \sqrt{2D(T)t},
\label{eq:four}
\end{equation}

where t is the annealing time. The height of the Bragg peak was
determined after subtracting the background due to Fresnel
reflectivity by multiplying the data by a factor of q$^{4}$, where
q is the momentum transfer. Fig.~\ref{fig:Fig11} shows an
evolution of the diffusion length as a function of annealing time
at 473 K. As can be seen from the figure, the diffusion lengths
below an annealing time of 600 s were found to increase much
faster as compared to later annealing times. Such behavior in
evolution of the diffusion length was also observed for
Fe$_{67}$Zr$_{33}$ amorphous sample~\cite{Gupta_PRB04} and is a
direct consequence of structural relaxation in the
structures.~\cite{Loirat_JNCS00,Gupta_PRB02}


It is interesting to see that for the sample prepared without any
stress, the diffusion length increased much faster as compared to
the samples prepared with an applied stress. As it is evident from
the structural and magnetic measurements, samples prepared with an
applied stress exhibited a more relaxed state as compared to that
obtained without an applied stress. It is expected that structural
relaxation would be more dominant for the sample prepared without
an applied stress. The overall magnitude of the diffusion length
follows the strength of applied stress and the degree of
relaxation is proportional.

\subsection {{\label{sec:RnD:SelfDiff:Temp}}Self-Diffusion Measurements - Temperature Dependence}

In order to measure the activation energy for diffusion, the
samples were annealed in the temperature range of 413-533 K with a
step of 40 K. As can be seen from the fig.~\ref{fig:Fig11}, after
an annealing time of 1800 s, in all cases, the fast relaxation
process was almost completed, therefore for the calculation of the
activation energy of the system all the samples were annealed for
1800 s at the above mentioned temperatures. It may be noted that
annealing for 1800 s may not produce a fully relaxed state of the
structure, even though for a comparison of diffusivities for the
samples prepared with different applied stress, the time for
diffusion annealing was kept constant.

\begin{figure}
\includegraphics[width=70mm,height=60mm]{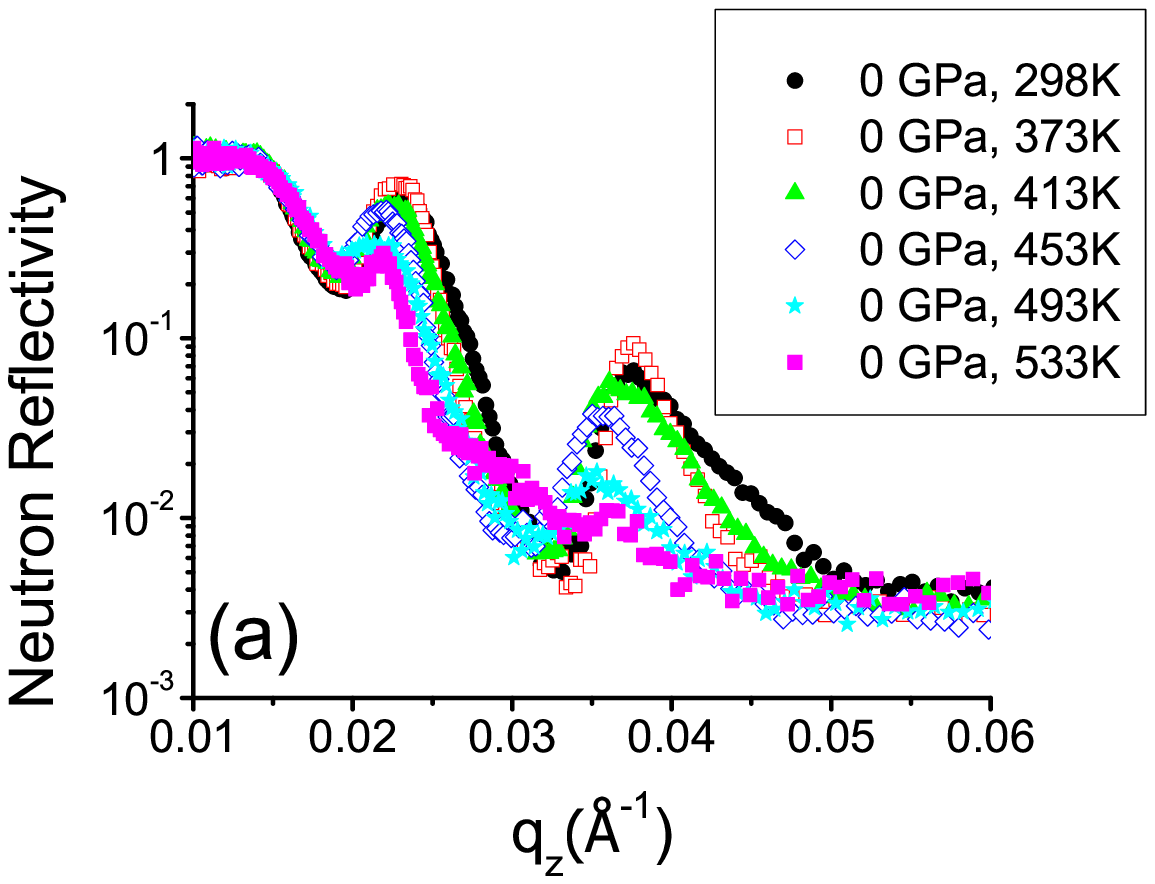} \vspace{-5mm}
\includegraphics[width=70mm,height=60mm]{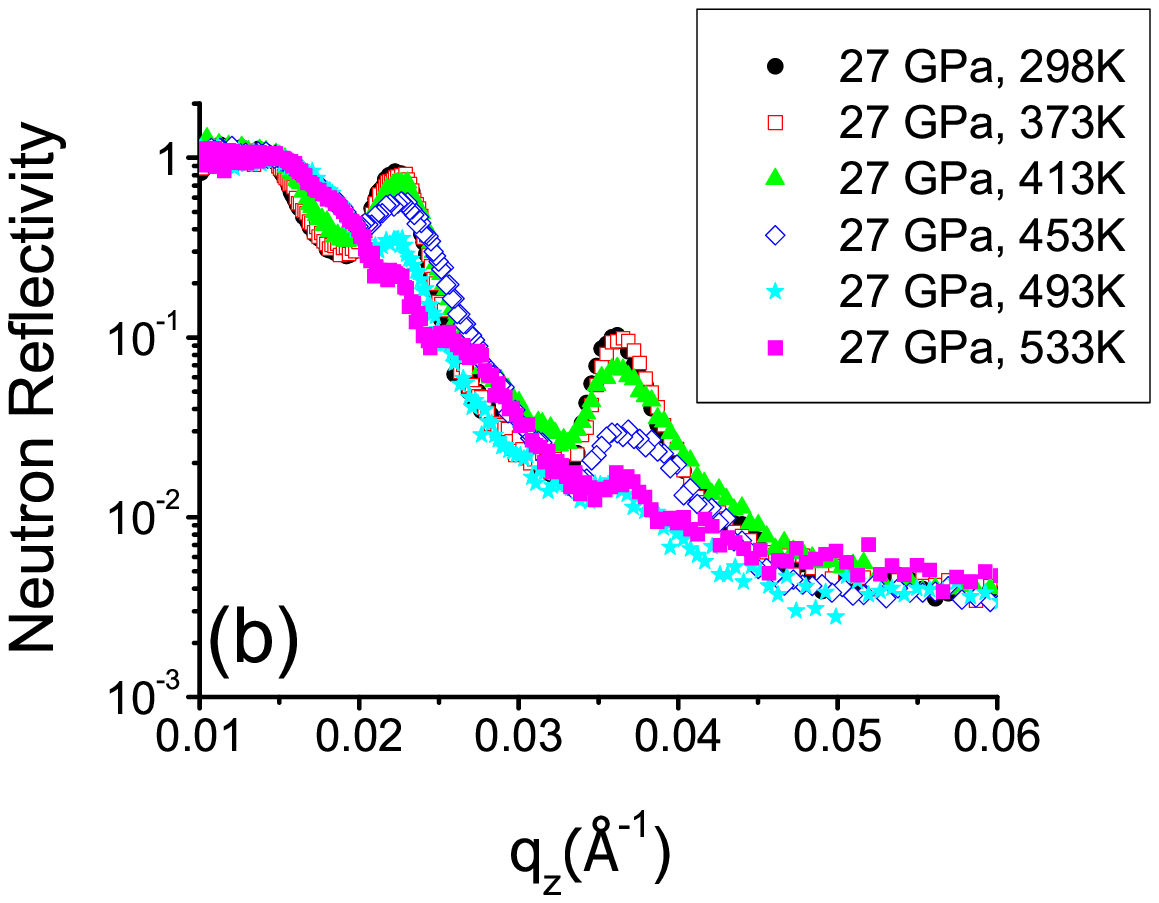} \vspace{-5mm}
\includegraphics[width=70mm,height=60mm]{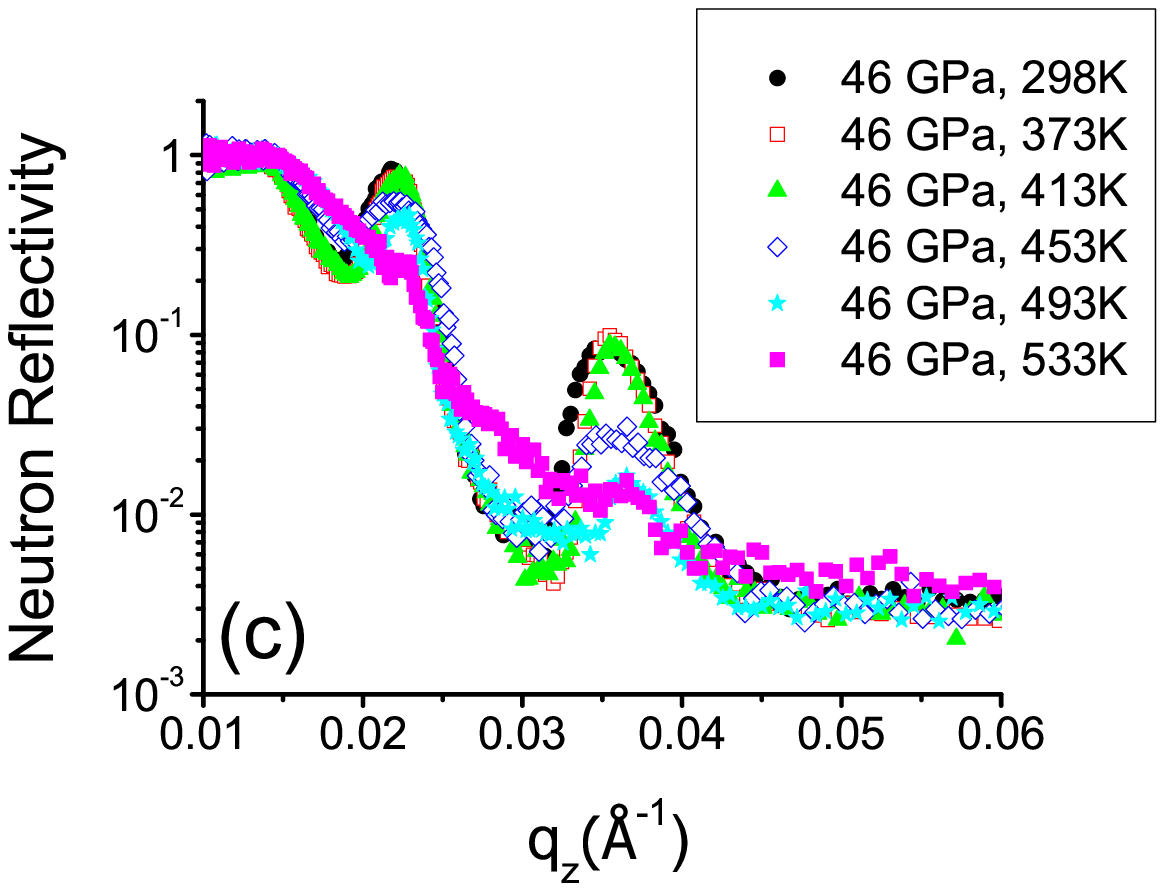}
\caption{\label{fig:Fig12:Bragg Decay 0,27,46GPa} (color online)
Decay of the Bragg peak intensity as a function of annealing
temperature for the sample prepared without an applied stress (a)
with an applied stress of 27 GPa (b) and with an applied stress of
46 GPa (c).}
\end{figure}

\begin{figure}
\includegraphics[width=65mm,height=55mm]{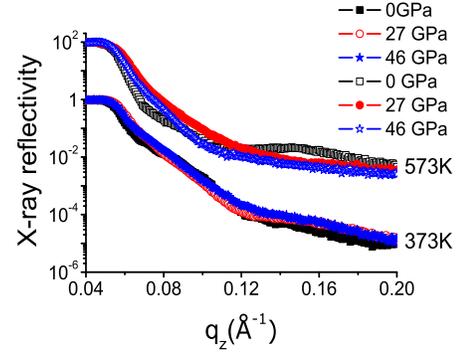}
\caption{\label{fig:Fig13} (color online)X-ray reflectivity
pattern of the Si/[$^{natrural}$Fe$_{75}$Zr$_{25}$ (25$\pm$1
nm)/$^{57}$Fe$_{75}$Zr$_{25}$(12$\pm$1 nm)]$_{10}$ multilayers
prepared with and without applied stress after annealing at 373
and 473 K }
\end{figure}

\begin{figure} \vspace{-5mm}
\includegraphics[width=70mm,height=60mm]{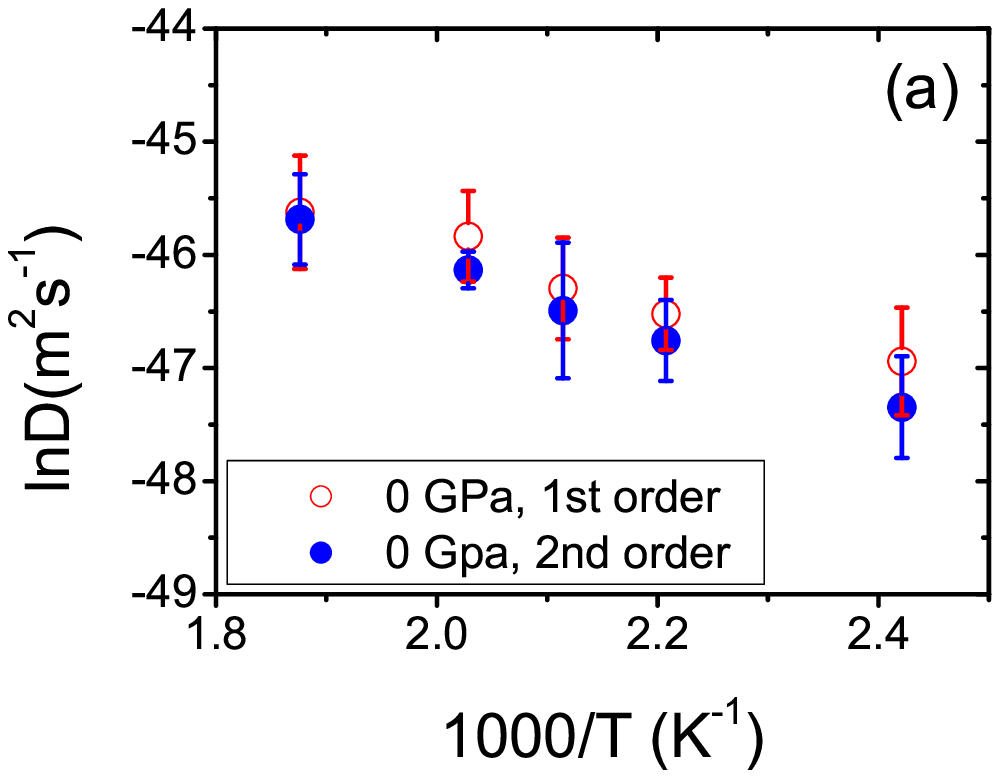} \vspace{-3mm}
\includegraphics[width=70mm,height=60mm]{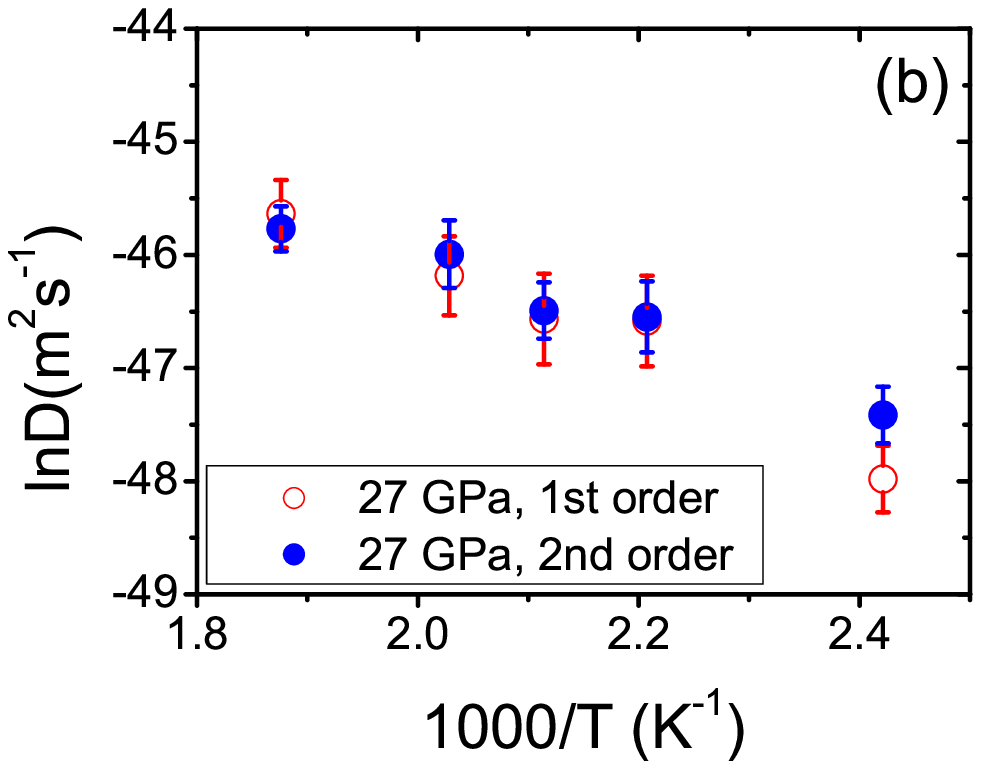} \vspace{-3mm}
\includegraphics[width=70mm,height=60mm]{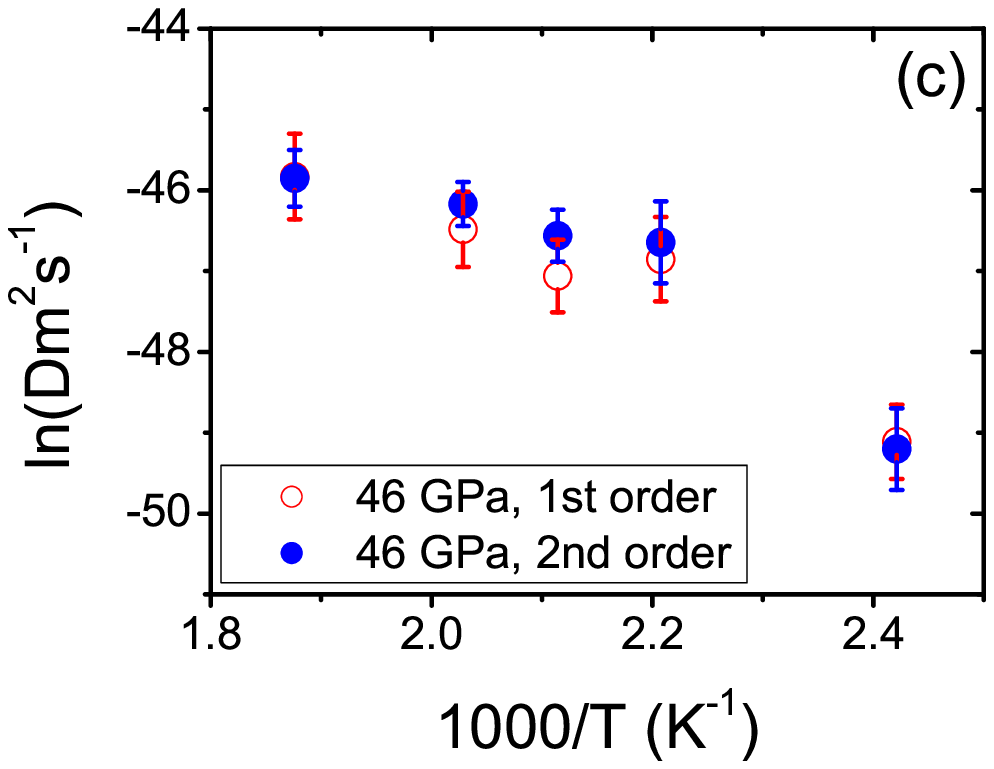}
\caption{\label{fig:Fig14} (color online) Arrhenius behavior of
the diffusivity obtained from the 1st and 2nd order Bragg peaks
for samples prepared with an applied stress of 0 GPa (a), 27 GPa
(b) and 46 GPa (c).}
\end{figure}

Fig.~\ref{fig:Fig12:Bragg Decay 0,27,46GPa} shows the neutron
reflectivity pattern obtained after annealing at different
temperatures. Again in order to minimize a possible fluctuation in
the temperature and annealing conditions, all the three samples
were annealed simultaneously in the furnace. For a comparison, the
neutron reflectivity pattern in the as-prepared state is also
shown in the figure. As evident from XRD and CEMS measurements,
after annealing the samples undergo from the amorphous to the
nano-composite state, the intensity at the Bragg peaks increases
marginally in the neutron reflectivity patterns. In a previous
study, it was observed that at the event of primary
crystallization in the amorphous Fe$_{67}$Zr$_{33}$
alloy~\cite{Gupta_PRB04} the neutron reflectivity pattern of an
[$^{natural}$Fe$_{67}$Zr$_{33}$(9 nm)\ $^{57}$Fe$_{67}$Zr$_{33}$(5
nm)]$_{20}$ multilayer, showed an increase in the intensity at the
Bragg peak by a factor as high as 10, as compared to the
as-deposited sample. Also the x-ray reflectivity pattern showed an
appearance of a Bragg peak due to crystallization accompanied by a
phase separation in the alloy.~\cite{Gupta_PRB04} In the present
case however, after annealing at 373 K, the amorphous phase
nanocrystallize, but the intensity at the Bragg peak increases
only marginally ($<$10\%). Also as shown in the
fig.~\ref{fig:Fig13}, no Bragg peak or structure due to a chemical
period appeared up to 573 K in the XRR pattern. This indicates
that the primary crystallization behavior of Fe$_{75}$Zr$_{25}$
alloy is different as compared to that of previously studied
Fe$_{67}$Zr$_{33}$ alloy.~\cite{Gupta_PRB04} However, since the
matrix obtained after nanocrystallization showed no further
significant changes between the temperature range 373-575 K (see
also fig.~\ref{fig:Fig8}), it is expected that the diffusion
process would be not interfered by structural changes. As shown in
fig.~\ref{fig:Fig12:Bragg Decay 0,27,46GPa}, the intensity at both
the Bragg peaks decreases with increase in annealing temperature
and after annealing at 533 K, the Bragg peak intensity almost
vanishes. This indicates that after annealing at 533 K, both the
natural and $^{57}$Fe are layers almost completely diffused. With
the procedure discussed in the previous section, the diffusivity
at each temperature was obtained. Fig.~\ref{fig:Fig14} shows a
plot of diffusivities obtained with both the Bragg peaks for the
three samples. As can be seen from the figure, both the Bragg
peaks yield similar diffusivities within the experimental errors.
The error bars in the present case are basically representing the
errors in determining the height of the Bragg peaks obtained from
a peak fitting procedure.

\begin{figure}
\includegraphics[width=75mm,height=70mm]{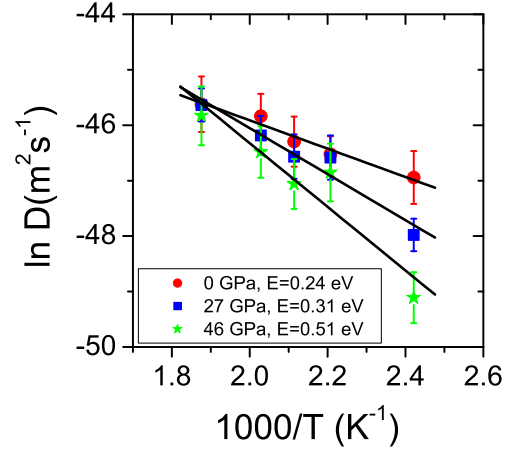} \vspace{-3mm}
\caption{\label{fig:Fig15} (color online) Activation energy and
pre-exponential factor for diffusion as a function of applied
stress in Si/[$^{natrural}$Fe$_{75}$Zr$_{25}$ (25$\pm$1
nm)/$^{57}$Fe$_{75}$Zr$_{25}$(12$\pm$1 nm)]$_{10}$ multilayer. The
data corresponds to the diffusivity obtained from the 1st order
Bragg peak. The sample prepared with the highest applied gives and
activation energy more than twice as compared with the sample
prepared without an applied stress. The detailed values of
activation energy and the pre-exponential factors are given in
table~\ref{Tab:Table2}.}
\end{figure}

\begin{table}
\caption{\label{Tab:Table2} Activation energy and the
pre-exponential factor for iron self-diffusion in nanocrystalline
Si/[$^{natrural}$Fe$_{75}$Zr$_{25}$ (25$\pm$1
nm)/$^{57}$Fe$_{75}$Zr$_{25}$(12$\pm$1 nm)]$_{10}$ multilayers as
a function of applied stress during sample preparation. Both the
activation energy and the pre-factor represent the statistical
averaged values obtained from the 1st and 2nd order Bragg peaks.}
\begin{ruledtabular}
\begin{tabular}{cccc}
Sample                        &Sample & Activation & Prefactor\\
                        & condition  & energy & (D$_{0}$, m$^{2}$s$^{-1}$)\\
                        &   & (E, eV) &  \\
\hline

& & & \\
nano.-Fe$_{75}$Zr$_{25}$      & 0 GPa            & 0.24$\pm$0.05       & 3$\times10^{-18\pm1}$  \\

nano.-Fe$_{75}$Zr$_{25}$      & 27 GPa           & 0.31$\pm$0.05      & 1$\times10^{-17\pm1}$ \\

nano.-Fe$_{75}$Zr$_{25}$      & 46 GPa           & 0.51$\pm$0.05       & 1$\times10^{-15\pm1}$ \\

amorphous-Fe$_{67}$Zr$_{33}$  & 0 GPa            & 0.38$\pm$0.05      & 3$\times10^{-18\pm1}$ \\
\end{tabular}
\end{ruledtabular}
\end{table}

The values for the diffusivity obtained for the three samples at
the abovementioned temperatures (along with the separately
annealed samples at 473 K), could be fitted to the relation,
$lnD=lnD_{0}-(E/k_{B}T)$, where $D_{0}$, $E$ and $T$ are the
pre-exponential factor, the activation energy and the annealing
temperature respectively and $k_{B}$ is the Boltzmann constant. In
all the three cases the logarithm of diffusivity follows Arrhenius
behavior and accordingly, the activation energy and the
pre-exponential factor for iron self-diffusion in
Fe$_{75}$Zr$_{25}$ alloy was obtained. The observed values of both
the E and D$_{0}$ are given in table~\ref{Tab:Table2} along with
the values obtained for amorphous Fe$_{67}$Zr$_{33}$ alloy.
Fig.~\ref{fig:Fig15} shows a plot of diffusivities obtained from
the 1st order Bragg peak for the sample at 0, 27 and 46 GPa. The
activation energy was found to increase with an increase in the
strength of applied compressive stress (a steeper slope was
observed with an increase in the applied stress). The result gives
a clear indication that diffusivity for the sample prepared with
applied stress is much slower as compared with that prepared
without an applied stress. This result also supports the time
dependence of the diffusivity as shown in fig.~\ref{fig:Fig11}.

In an earlier study Klugkist et
al~\cite{Klugkist_PRL98a,Klugkist_PRL98b} studied Co and Zr
self-diffusion in amorphous CoZr alloy using Radioactive tracer
method as a function of pressure and temperature dependence. It
was found that the pressure dependence for Co self-diffusion is
extremely small while for Zr self-diffusion it is of the order of
one activation volume. On the basis of obtained results it was
concluded that Zr diffuses via thermal defects, whereas vacancy
like thermal defects can be ruled out for Co self-diffusion.
However, our results clearly indicates a decrease in diffusivity
with an increase in applied stress. Here we would like to point
out that a direct comparison between the studies performed earlier
~\cite{Klugkist_PRL98a,Klugkist_PRL98b} with that of our results
could not be made. The following points are important to
understand our results (i) The neutron reflectometry technique
offers a depth resolution in sub nano meter range it is possible
to measure initial stage of self-diffusion in a alloy unlike
conventional techniques e.g. SIMS or radioactive tracer method
(ii) As mentioned already that in our case, the alloy could have
not attained a fully relaxes state as pre-annealing time was very
short compared with that in literature. Combining the above
mentioned points, it is not surprising that there is a strong
dependence of self-diffusion of Fe on applied stress, which points
out that in the initial state the diffusion mechanism could be
different.

Comparing the diffusivity for the sample prepared at 0 GPa with
that of amorphous Fe$_{67}$Zr$_{33}$ sample (also prepared at 0
GPa), the diffusivity in the nano-composite state is slightly
higher as compared to the amorphous sample. The activation energy
for the nanocomposite sample was lower by 0.14 eV, while the
pre-exponential factors were found to be exactly similar (see
table ~\ref{Tab:Table2}). An enhancement in diffusivity in the
nano-composite state is not unexpected due to presence of grains
and grain-boundaries (GB), while the amorphous phase is expected
to be free from grains and GBs. However the enhancement in
diffusivity in the present case is not as spectacular as observed
e.g. in the FINEMET type
nanocrystalline-Fe$_{73.5}$Si$_{13.5}$B$_{9}$Nb$_{3}$Cu$_{1}$ in
which the Fe self-diffusion showed a large enhancement over that
in the parent amorphous phase.~\cite{Tanimoto_NM99,Wurschum_PRL97}
It may be noted that in the present case the composition of the
nano-composite Fe$_{75}$Zr$_{25}$  alloy is also not similar to
the amorphous Fe$_{67}$Zr$_{33}$ alloy, therefore an enhancement
in diffusivity may also occur due to increased concentration of
Fe. In another study self-diffusion of iron was measured in the
parent amorphous and nano-composite Fe$_{85}$Zr$_{15}$  alloy thin
film produced by ion-beam sputtering. It was found that iron
self-diffusion in both amorphous and nano-composite state was
similar and found to occur exclusively through the GB regions
which were amorphous in nature.~\cite{Gupta_JNCS04} In the present
case as well the GBs in the nano-composite state are amorphous,
which might happen due to the fact that in the nano-composite
state, the structure consists of a mixture of nanocrystalline
grains of Fe and remaining amorphous phase. The nanocrystalline
grains of Fe would be surrounded by amorphous GBs and in such a
situation a percolating path between the nanocrystals may not
establish and diffusivity in the nano-composite phase would be
similar to that in the amorphous state.

On the other hand, the activation energy obtained for the sample
prepared with highest applied stress was found to be larger
(slower diffusivity) as compared to the amorphous
Fe$_{67}$Zr$_{33}$ sample. This is somewhat surprising as the
sample prepared even at the highest stress is also in the
nano-composite state. In case the diffusion mechanism is dominated
by grains and GBs the effective applied stress should result in an
enhancement of the diffusivity.~\cite{Crosby_2003} As discussed
earlier, an applied compressive stress produced a more relaxed
state of the sample as compared to samples prepared without
stress. If the diffusion mechanism is dominated by a somewhat
collective type migration of atoms both in amorphous and
nano-composite case, annihilation of free volume would result into
a diffusion mechanism involving a small group of atoms.

\section{\label{sec:Conclusion} CONCLUSIONS}
In the present work, the effect of compressive stress on
self-diffusion of iron in chemically homogenous multilayers of
FeZr/$^{57}$FeZr was investigated. It was found that samples in
the as-prepared state were amorphous and undergo primary
crystallization when annealed at 373 K. The diffusion measurements
were performed in the nano-composite state and it was observed
that with an increase in the strength of applied stress, the
diffusivity decreases as compared to the sample prepared without
an applied stress. An applied compressive stress on to the
multilayer produced a more relaxed state of the sample as seen
from XRD. A diffusion mechanism involving a small group of atoms
explains the observed diffusivity in the chemically homogeneous
multilayers.

\section{\label{sec:Acknowledgement} ACKNOWLEDGEMENT}
Authors would like acknowledge: A. Foelske, General Energy
Research, Paul Scherrer Institute, for providing help in x-ray
photoelectron spectroscopy measurements, K. Conder, P. Keller, and
M. Horisberger, Laboratory for Neutron Scattering, Paul Scherrer
Institute for providing help in DSC measurements, manufacturing of
3-point Si wafer bending machine, assisting in thin film
deposition, respectively. This work was performed at the Swiss
Spallation Neutron Source, Paul Scherrer Institute, Villigen,
Switzerland.

\bibliography{Gupta_FeZr_Stress}

\end{document}